\documentclass[jkps,twocolumn,fleqn,showpacs,showkeys]{revtex4}
\usepackage{graphicx}
\usepackage{amssymb}
\usepackage{amsmath}
\usepackage{bm}

\usepackage{color}

\begin{document}
\setcounter{page}{0}
\title[]{

$1/(N-1)$ Expansion Approach to   
Full-counting Statistics for the SU($N$) Anderson Model
}
\author{Akira \surname{Oguri}}

\affiliation{Department of Physics, Osaka City University, Sumiyoshi-ku, 
Osaka, Japan}

\author{Rui \surname{Sakano}}
\affiliation{Institute for Solid State Physics, University of Tokyo, Kashiwa, Chiba, Japan}

\date{\today}

\begin{abstract}

We apply a recently developed $1/(N-1)$ expansion to 
the full-counting statistics for 
the  $N$-fold degenerate Anderson impurity model in the Kondo regime.
This approach is based on the perturbation theory in 
the Coulomb interaction $U$ and is 
different from the conventional large-$N$ theories, 
such as the usual $1/N$ expansion and non-crossing approximation.
We have confirmed that the calculations carried out up to order $1/(N-1)^2$  
 agree closely with those of the numerical renormalization group  
at $N=4$, where the degeneracy is still not so large.
This ensures the applicability of our approach for $N \geq  4$. 
We present the results of the cumulants of   
the probability distribution function 
for a nonequilibrium current through a quantum dot
in the particle-hole symmetric case. 
\end{abstract}

\pacs{72.15.Qm, 73.63.Kv, 75.20.Hr}

\keywords{Kondo effect, Quantum dot,  
Orbital degeneracy, Nonequilibrium current}

\maketitle

\section{Introduction}

The universal Kondo behavior in quantum dots has been studied 
intensively for a nonequilibrium current 
\cite{Grobis,ScottNatelson,KNG,ao2001,FujiiUeda,HBA} 
and shot noise 
\cite{Heiblum,Kobayashi,GogolinKomnik,GogolinKomnikPRL,Golub,Sela2009,Fujii2010}.
The effects of orbital degeneracy on the universality have 
also been  a subject of current  interest 
\cite{Delattret,Mora2009,Sakano,SakanoFCS,SakanoFCS2}. 

The essential features of the low-energy properties of orbital systems 
may be deduced from the SU($N$) Anderson model, 
 for which $N=2$ corresponds to the spin degeneracy.
The exact numerical renormalization group (NRG) approach  \cite{KWW} 
has been applied to this model for small degeneracies  $N \leq 4$ 
\cite{Izumida}.
For $N>4$, the conventional large-$N$ theories, 
such as the $1/N$ expansion and the non-crossing approximation 
\cite{Coleman,Bickers}, are applicable to a wide energy scale 
except for the low-energy Fermi-liquid regime.
Therefore,  at low energies, alternative approaches are needed 
to explore the universal behavior for $N>4$. 

Recently,
we proposed a different type of large-$N$ theory,
by using a scaling that takes $u =(N-1)U$ as an independent variable 
instead of the bare Coulomb interaction $U$ 
\cite{AoSakanoFujii,Ao2012}.
The factor $N-1$  represents the number of interacting orbitals, 
excluding the one prohibited by the Pauli principle. 
With this scaling, the perturbation series in $U$  
can be reorganized as an expansion in powers of $1/(N-1)$. 
To leading order in $1/(N-1)$, it describes the Hartree-Fock 
random phase approximation (HF-RPA). The higher-order 
corrections  systematically describe    
the fluctuations beyond the HF-RPA. 
As the unperturbed Hamiltonian includes the 
tunneling matrix element 
between the impurity and the conduction bands, 
this approach naturally describes the Fermi-liquid state.

In this report, we discuss dependences 
of the renormalized parameters on the degeneracy $N$,  
carrying out the calculations up to order $1/(N-1)^2$.
Furthermore, we apply this approach to  
the full-counting statistics \cite{LevitovReznikov,Bagrets,Esposito}
for the nonequilibrium current distribution 
through quantum dots in the Kondo regime.

\section{Model and Formulation}
 \label{sec:formulation}

We consider the SU($N$) impurity Anderson model    
with a finite interaction $U$  connected to two leads  
($\nu=L,\,R$):  
${\cal H} =   {\cal H}_0 + {\cal H}_U$,  
\begin{align}
&
\!\!\!\!\!\!\!\!\!\!
{\cal H}_0 =  
\sum_{m=1}^N \xi_d^{}
 \,d_{m}^{\dagger} d_{m}^{} 
+ \sum_{\nu=L,R}
\sum_{m=1}^N 
\int_{-D}^D  \! d\epsilon\,  \epsilon\, 
 c^{\dagger}_{\epsilon \nu m} c_{\epsilon \nu m}^{}
 \nonumber \\
&  \ \     + 
\sum_{\nu=L,R}\,
\sum_{m=1}^N \,
 v_{\nu}^{} \,\Bigl(
d_{m}^{\dagger} \psi^{}_{\nu m} 
+ \psi^{\dagger}_{\nu m} d_{m}^{} 
\Bigr) \;,
\\
&
\!\!\!\!\!\!\!\!\!\!
{\cal H}_U 
=  \,  \sum_{m\neq m'} \frac{U}{2}
\left(  n_{dm}^{}-\frac{1}{2} \right)
\left(  n_{dm'}^{} -\frac{1}{2} \right)\;.
\label{eq:AndersonHamiltonian}
\end{align}
Here,  $\xi_d = \epsilon_{d}^{} + (N-1)U/2$,
$d_m^{\dagger}$ is the creation operator for an electron 
with energy $\epsilon_{d}$ 
and orbital $m$ ($=1,2, \cdots, N$) 
in the impurity site, and 
$n_{dm}^{} =\, d_{m}^{\dagger} d_{m}^{}$. 
The operator $c_{\epsilon \nu m}^{\dagger}$ 
for a conduction electron in the lead $\nu$  
is normalized as    
$
\{ c^{\phantom{\dagger}}_{\epsilon \nu m}, 
c^{\dagger}_{\epsilon'\nu'm'} 
\} = \delta_{\nu\nu'}\,\delta_{mm'}   
\delta(\epsilon-\epsilon')$, 
and  $\psi^{}_{\nu m} =  \! \int_{-D}^D \! d\epsilon \sqrt{\rho} 
\, c^{\phantom{\dagger}}_{\epsilon \nu m}$.
 The hybridization energy scale is given by 
$\Delta\equiv \Gamma_L+\Gamma_R$, with  
 $\Gamma_{\nu} = \pi \rho\, v_{\nu}^2$  and $\rho=1/(2D)$.
%

We use the imaginary-frequency Green's function 
that is given, for $|\omega| \ll D$, by 
\begin{align}
G(i\omega) 
\,=\,  \frac{1}{i\omega - \xi_d^{} 
+ i \Delta \, \mathrm{sgn}\,\omega 
- \Sigma (i\omega)} \;.
\label{eq:G_dd}
\end{align}
Here, 
$\Sigma (i\omega)$ is the self-energy due to ${\cal H}_U$.
The ground-state average of the local charge, 
$\langle n_{dm} \rangle = \delta/\pi$,  
can be deduced from the phase shift  
$\delta \equiv  \cot^{-1} (E_d^*/\Delta)$  
with $E_d^* \equiv  \xi_{d}+\Sigma(0)$.
The renormalized parameters are defined by 
$
1/z \equiv
1 -  {\partial\Sigma(i\omega)}/{\partial (i\omega)}
|_{\omega=0}$, 
$\,\widetilde{\epsilon}_{d} \equiv   z\,E_{d}^*$, 
and 
$\widetilde{\Delta}  \equiv z \Delta$. 
Furthermore, 
the enhancement factors for 
the spin and charge  susceptibilities, 
$\widetilde{\chi}_{s}^{} \equiv
\widetilde{\chi}_{mm}^{} - 
\widetilde{\chi}_{mm'}^{}$ and 
$\widetilde{\chi}_{c}^{}  \equiv 
\widetilde{\chi}_{mm}^{} + (N-1)\, \widetilde{\chi}_{mm'}^{}$, 
can be expressed in terms of  
$z$ and the vertex function  
 $\Gamma_{mm';m'm}^{}(i \omega_1,i\omega_2;i\omega_3,i\omega_4)$ 
for $m\neq m'$ \cite{YY2,Yoshimori}:    
\begin{align}
\ \!\!\!\!\!\!\!\!\!\!\!\!\!\!
\widetilde{\chi}_{mm}^{} = 
\frac{1}{z} , 
\quad \ \ 
\widetilde{\chi}_{mm'}^{} =    -\,
\frac{\sin^2 \! \delta}{\pi\Delta}
\, \Gamma_{mm';m'm}^{}(0,0;0,0) ,
\label{eq:ward_fermi}
\end{align}
and  $\widetilde{U} 
\equiv z^2 \Gamma_{mm';m'm}^{}(0,0;0,0)$ for $m \neq m'$ 
represents the residual interaction between the quasi-particles.

We introduce a scaling  for the bare and the renormalized interactions 
with a factor $N-1$ of \cite{AoSakanoFujii,Ao2012}:
\begin{align} 
g \equiv \frac{(N-1)\,U}{\pi\Delta} \;,
\qquad \qquad 
\widetilde{g} \,\equiv\, 
\frac{ (N-1)\,\widetilde{U}}{\pi \widetilde{\Delta}} \;.
\end{align}
With these parameters, the Wilson ratio 
 $R \equiv z\, \widetilde{\chi}_{s}$ 
and that for the charge sector 
can be expressed in the form, 
\begin{align}
& \!\!\!\!\!\!
R  
 =  1+\frac{\widetilde{g}}{N-1} \,\sin^2 \! \delta 
,
\qquad \quad 
z \, \widetilde{\chi}_{c}
 =   1- \widetilde{g}\,\sin^2 \! \delta .
\label{eq:Wilson}
\end{align}
 One of the merits of this scaling is that  
the perturbation expansion with respect to ${\cal H}_U$   
can be classified according to the power of $1/(N-1)$,
taking  $\,g$ as an independent variable.  
In the present report,  
we consider the particle-hole symmetric case:     
 $\epsilon_d =-(N-1)U/2$ and $\delta = \pi/2$.
%

\begin{figure}[t]
 \leavevmode
\begin{minipage}{1\linewidth}
\includegraphics[width=0.75\linewidth]{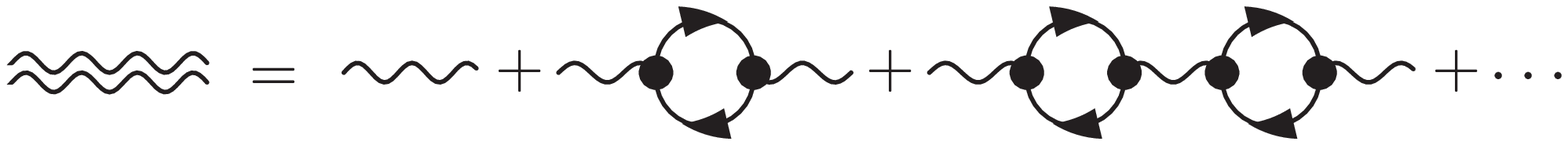}
\end{minipage}
\caption{The leading order diagrams in the $1/(N-1)$ expansion.
The wavy and the solid lines indicate 
 $U$ and 
 $G_0$, respectively.
The double wavy line 
represents $\mathcal{U}_\mathrm{bub}(i\omega)$. 
}
 \label{fig:vertex_rings}
\end{figure}

\begin{figure}[t]
\begin{minipage}{1\linewidth}
\includegraphics[width=0.24\linewidth]{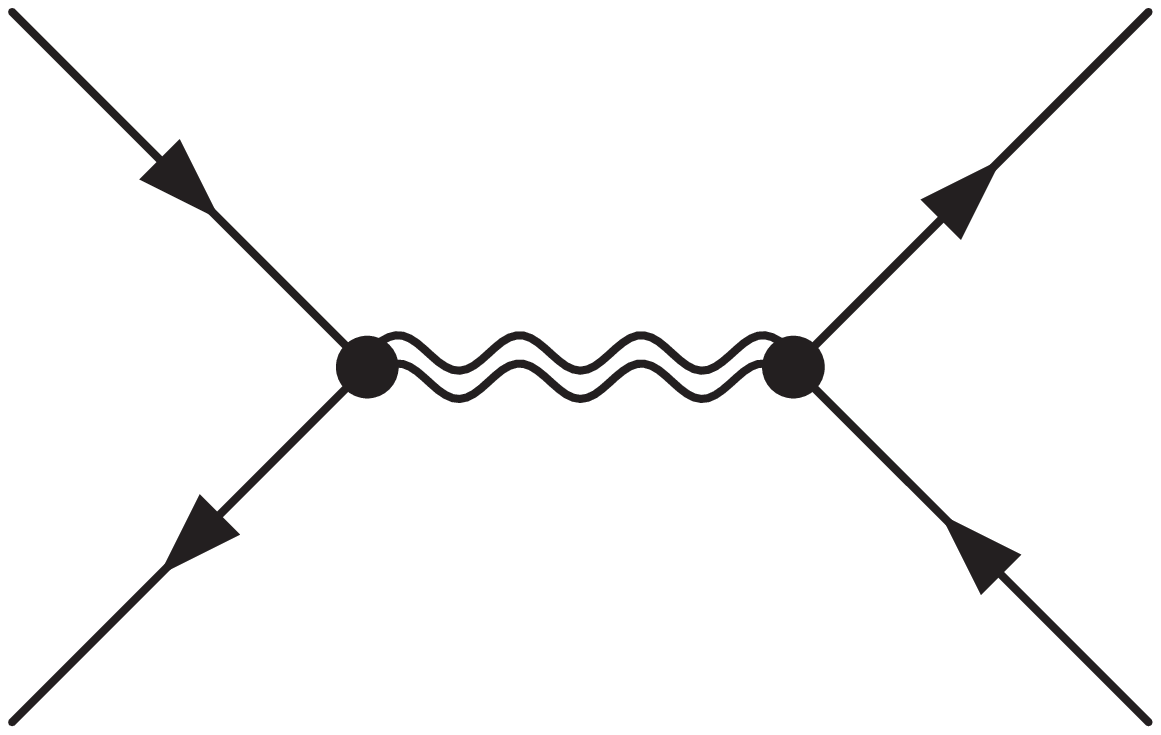}
\rule{0.15\linewidth}{0cm}
\includegraphics[width=0.35\linewidth]{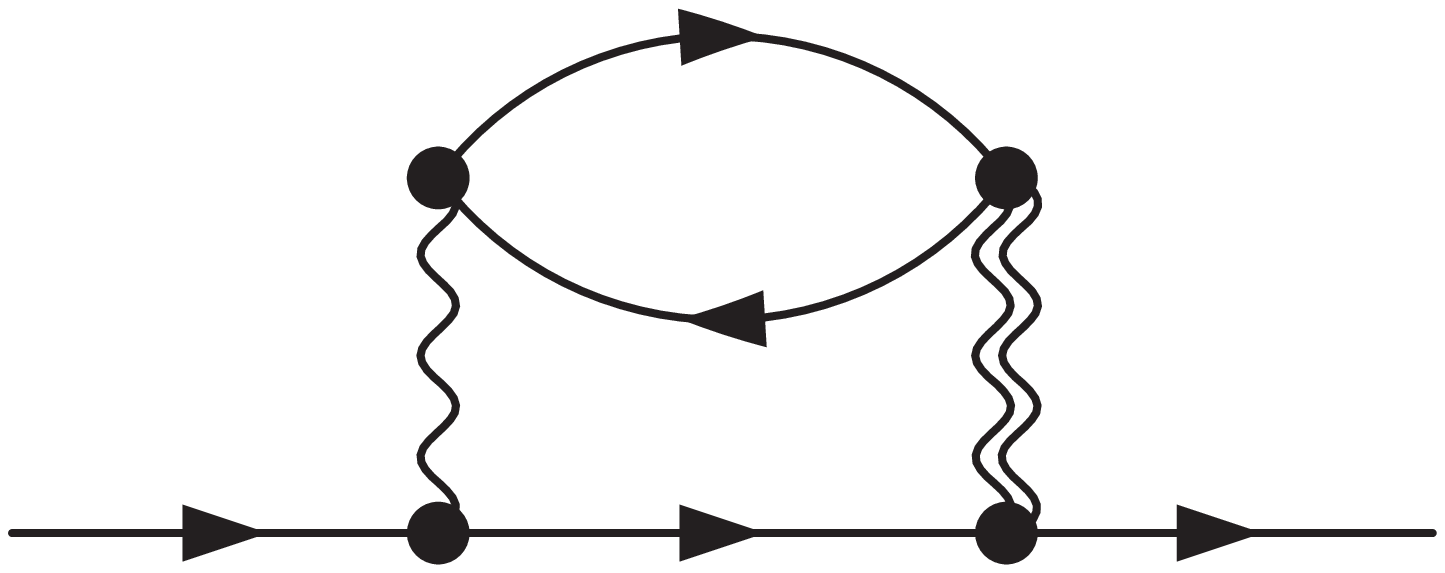}
\end{minipage}
\caption{
The order $1/(N-1)$ vertex and self-energy.  
}
 \label{fig:sg_rpa}
\end{figure}

\section{Expansion up to order $1/(N-1)^2$} 
\label{sec:1_(N-1)_expansion}

At  zero order with respect to $1/(N-1)$,  
the limit of $N \to \infty$ is taken at fixed $g$, 
and it gives the Hartree-Fock results. 
Specifically, at half-filling,  
the unperturbed Green's function is given by $G_0(i\omega) 
= \left[i\omega + i \Delta \,\mathrm{sgn}\, \omega\right]^{-1}$ 
as $\xi_d=0$.
%


The leading-order corrections in the $1/(N-1)$ expansion 
arise form a series of bubble diagrams, 
$\mathcal{U}_\mathrm{bub}(i\omega)$, 
 of the RPA typeas  shown in Fig.\ \ref{fig:vertex_rings}:  
\begin{align}
\mathcal{U}_\mathrm{bub}(i\omega) 
\, =&  \ \, 
U +  U \sum_{k=1}^{\infty} \, 
 \mathcal{A}_k\,\Bigl[- U\,\chi_0(i\omega) \Bigr]^k \;.  
\label{eq:U_bub_def}
\end{align}
Here, 
$\chi_0^{}(i\omega)  =    - 
\int
 \! \frac{d\omega'}{2\pi} 
G_0^{}(i\omega +i\omega') G_0^{}(i\omega')$, and    
the coefficient $\mathcal{A}_k
= (N-1)^k \, \sum_{p=0}^k \, \left[{-1}/{(N-1)} \right]^p 
$ 
arises from the summation over the orbital indices 
for a series of $k$ fermion loops. 
%
%
Correspondingly, 
the order $1/(N-1)$  contributions of $\Gamma_{mm';m'm}^{}(0,0;0,0)$ 
arise from the first diagram in Fig.~\ref{fig:sg_rpa},
and give the leading-order correction to the renormalized coupling as  
\begin{align}
\widetilde{g} \,=\,   \frac{g}{1+g} 
\,+\,O\!\left(\!\frac{1}{N-1}\!\right) .
\label{eq:g_inf}
\end{align}
This correction determines the Wilson ratio $R$ to order $1/(N-1)$  
through Eq.\ \eqref{eq:Wilson}.  
Similarly, 
the  order  $1/(N-1)$ self-energy 
arises from the second diagram in Fig.\ \ref{fig:sg_rpa}. 

\begin{figure}[t]
 \leavevmode
\begin{minipage}{1\linewidth}
\includegraphics[width=0.27\linewidth]{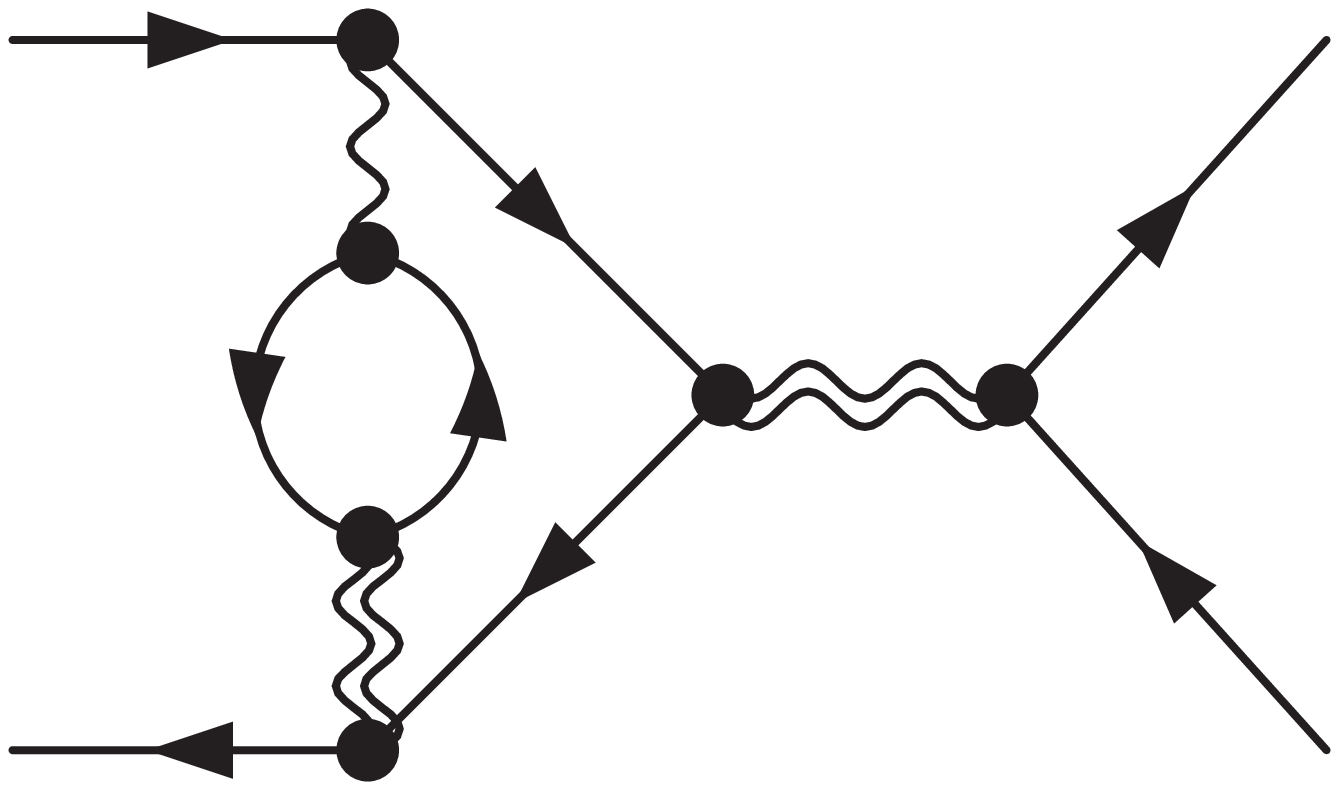}
 \rule{0.1\linewidth}{0cm}
\includegraphics[width=0.27\linewidth]{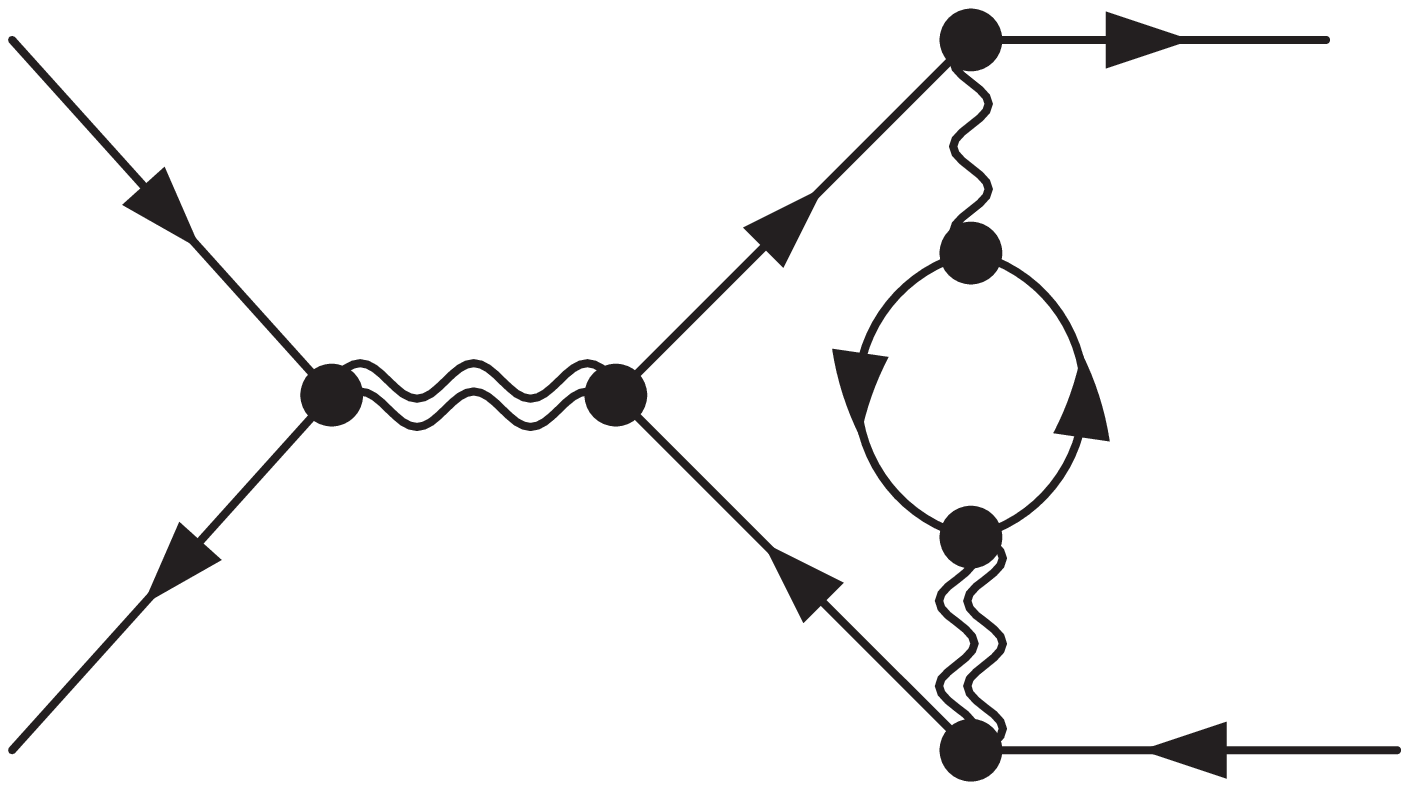}
\end{minipage}
 \rule{0cm}{0.15cm}
 \begin{minipage}{1\linewidth}
 \includegraphics[width=0.27\linewidth]{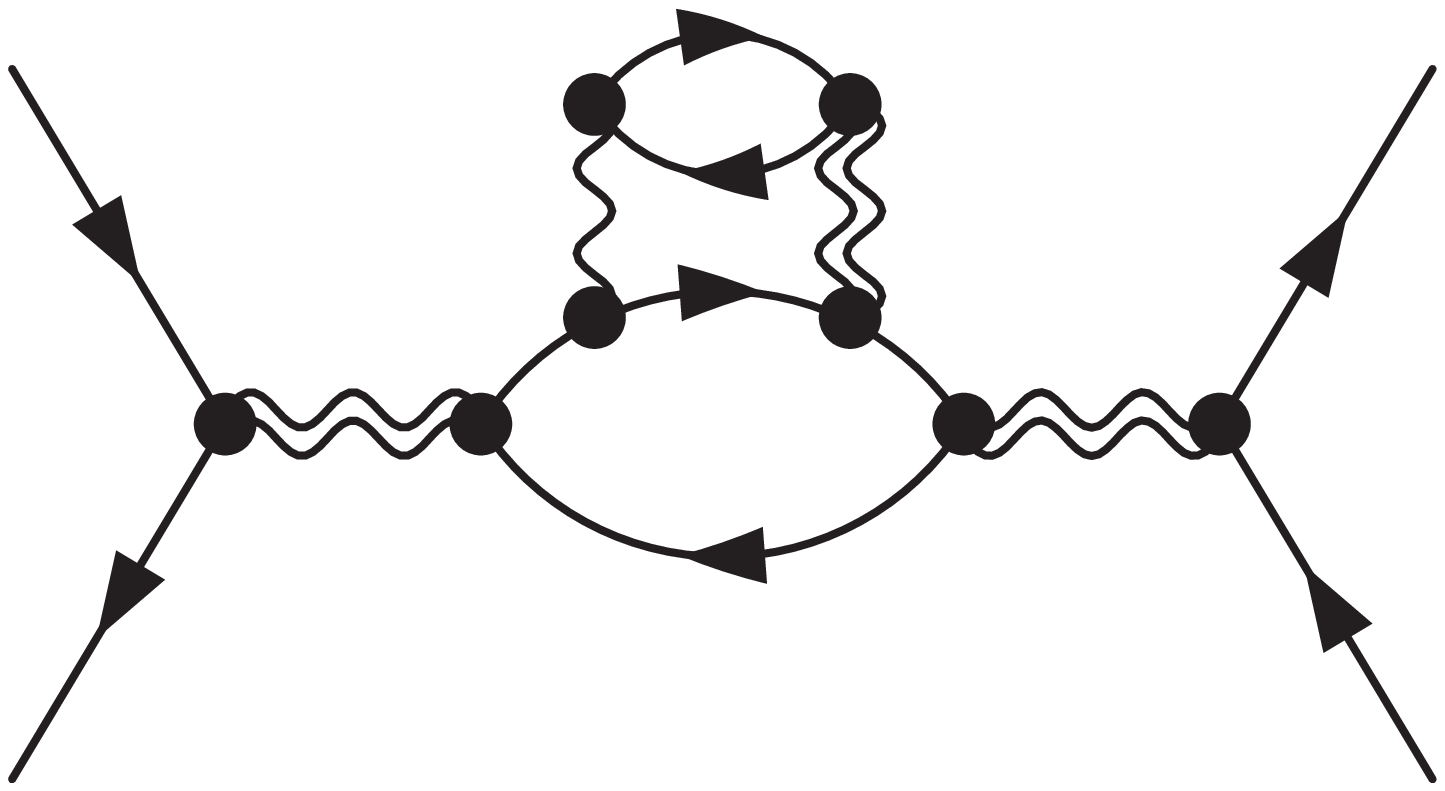}
 \rule{0.04\linewidth}{0cm}
 \includegraphics[width=0.27\linewidth]{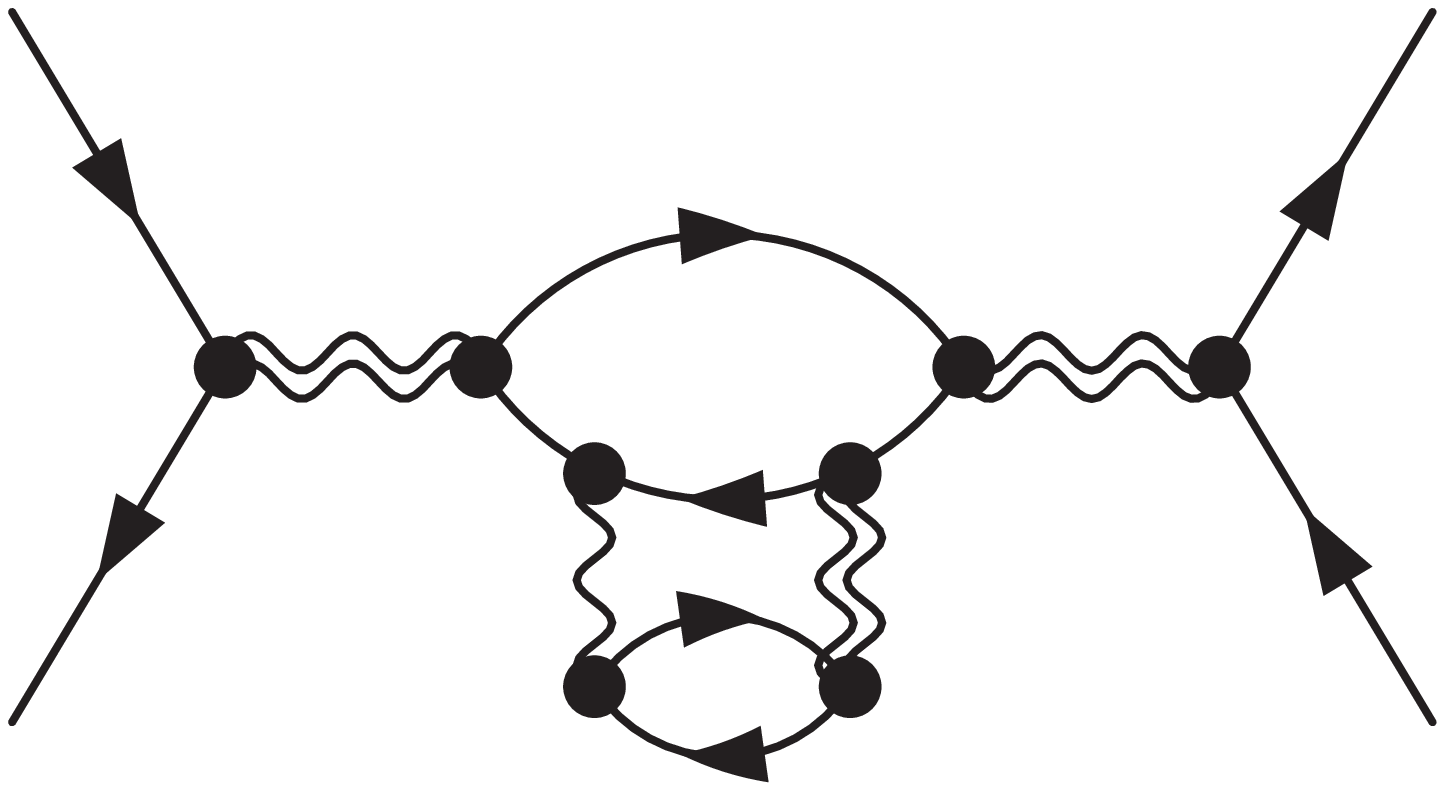}
 \rule{0.04\linewidth}{0cm}
 \includegraphics[width=0.27\linewidth]{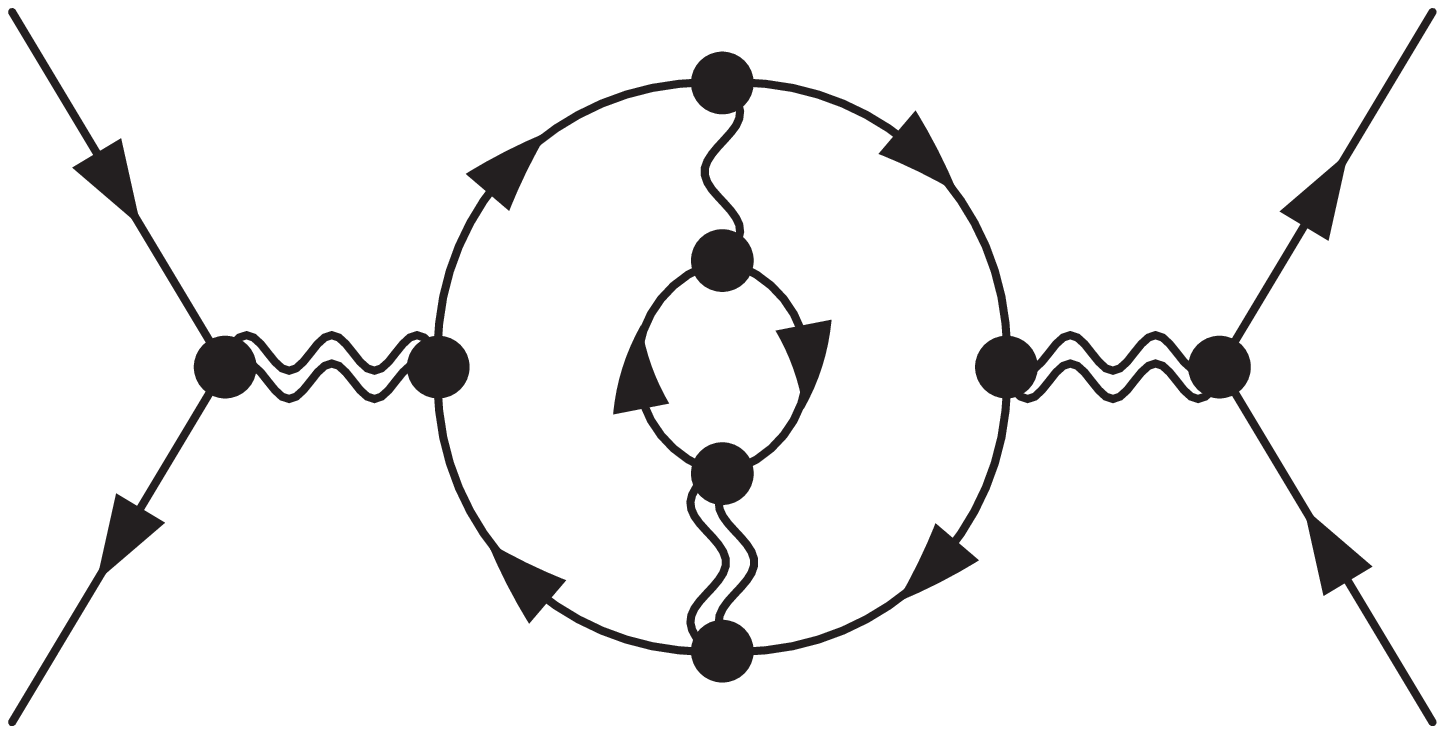}
 \end{minipage}
 \caption{
The order $1/(N-1)^2$ diagrams for the vertex function 
$\Gamma_{mm';m'm}^{}(0,0;0,0)$ for $m \neq m'$ 
at half-filling  $\,\xi_d=0$.
}
 \label{fig:vertex_rpa}
\end{figure}

\begin{figure}[t]
 \leavevmode
\begin{minipage}{1\linewidth}
\includegraphics[width=0.23\linewidth]{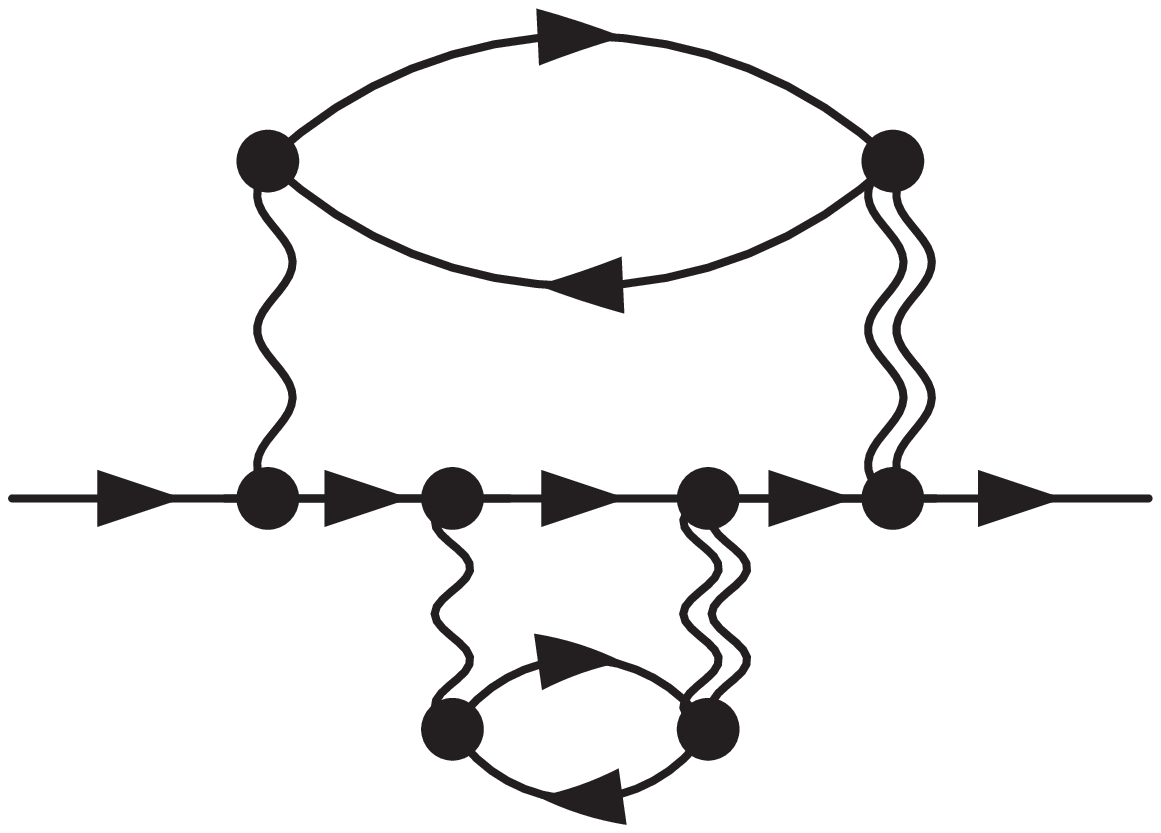}
 \rule{0.1\linewidth}{0cm}
\raisebox{-0.3cm}{
\includegraphics[width=0.25\linewidth]{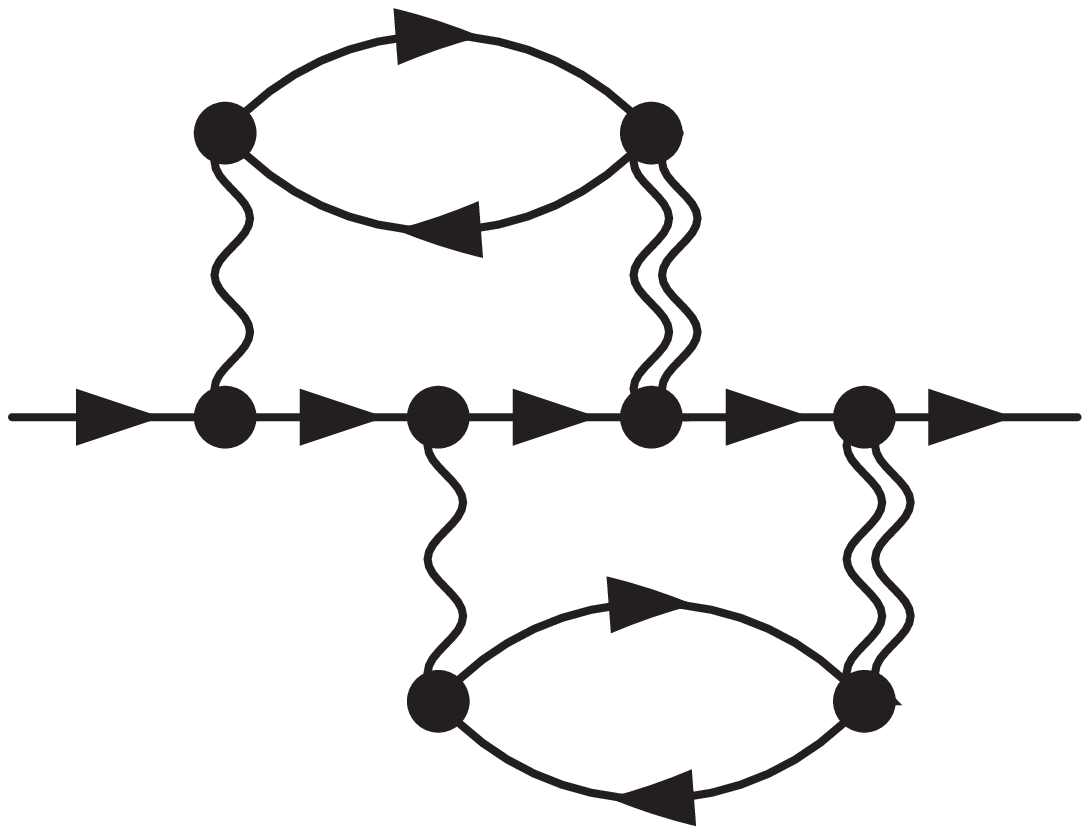}
}
\end{minipage}

\vspace{-0.15cm}

 \begin{minipage}{1\linewidth}
 \includegraphics[width=0.27\linewidth]{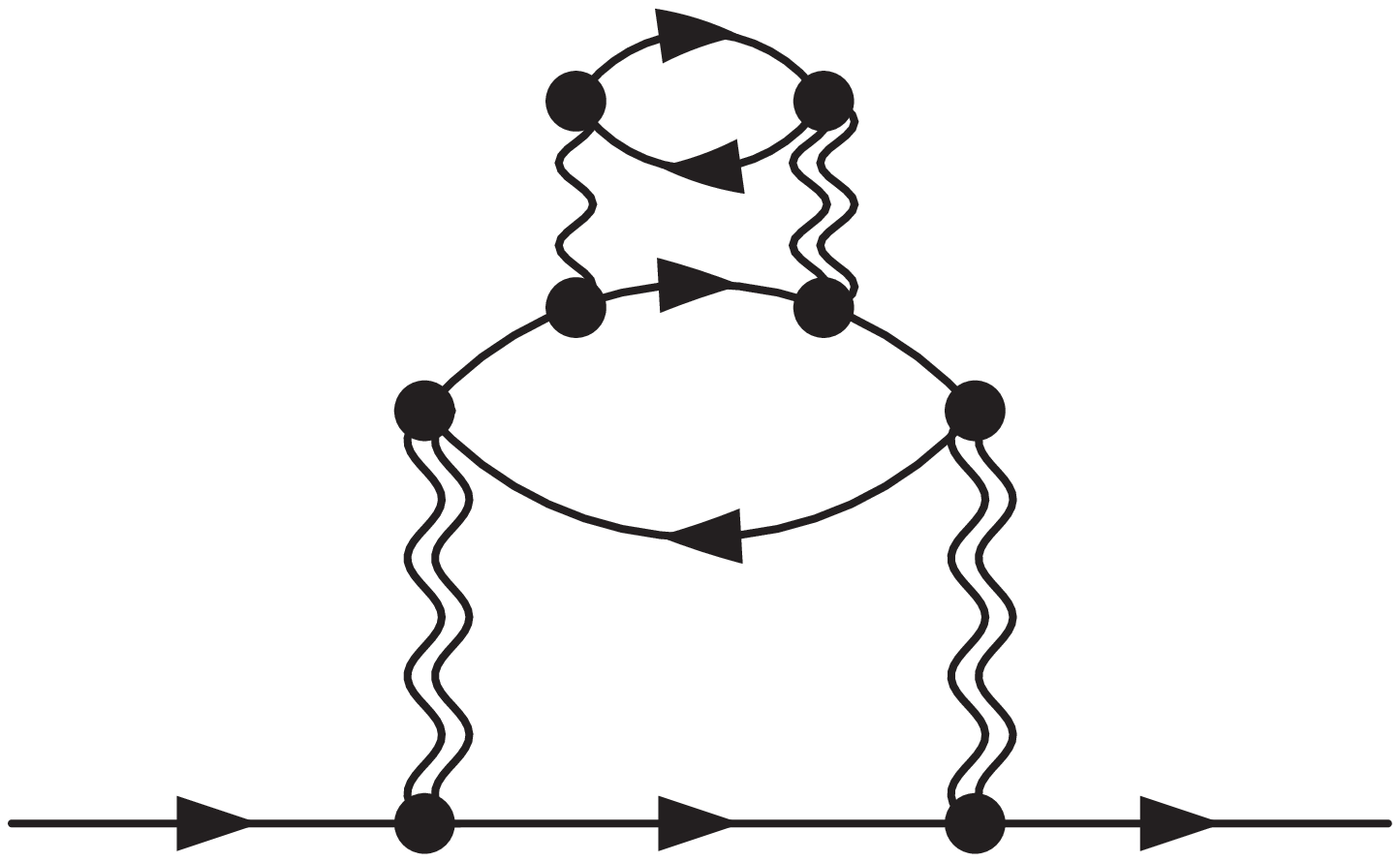}
 \rule{0.04\linewidth}{0cm}
 \includegraphics[width=0.27\linewidth]{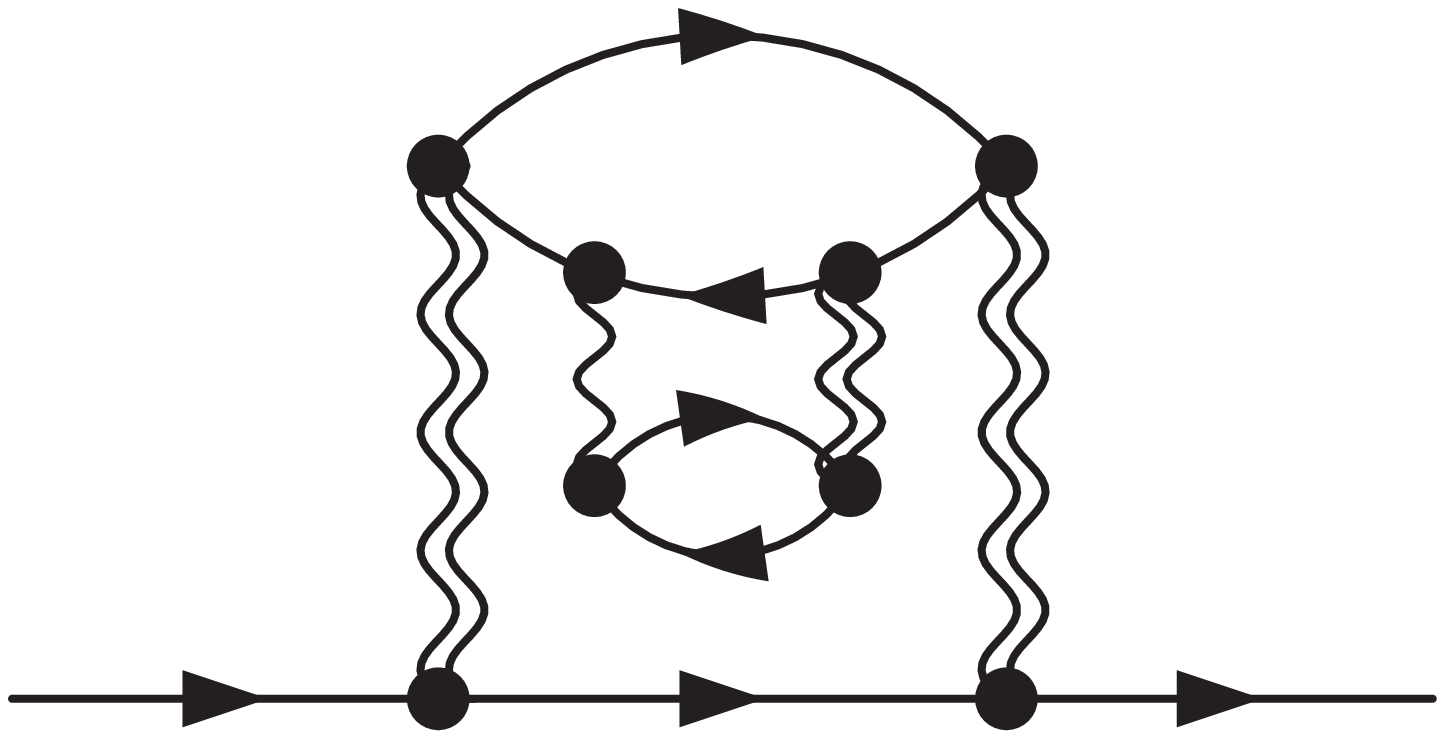}
 \rule{0.04\linewidth}{0cm}
 \includegraphics[width=0.27\linewidth]{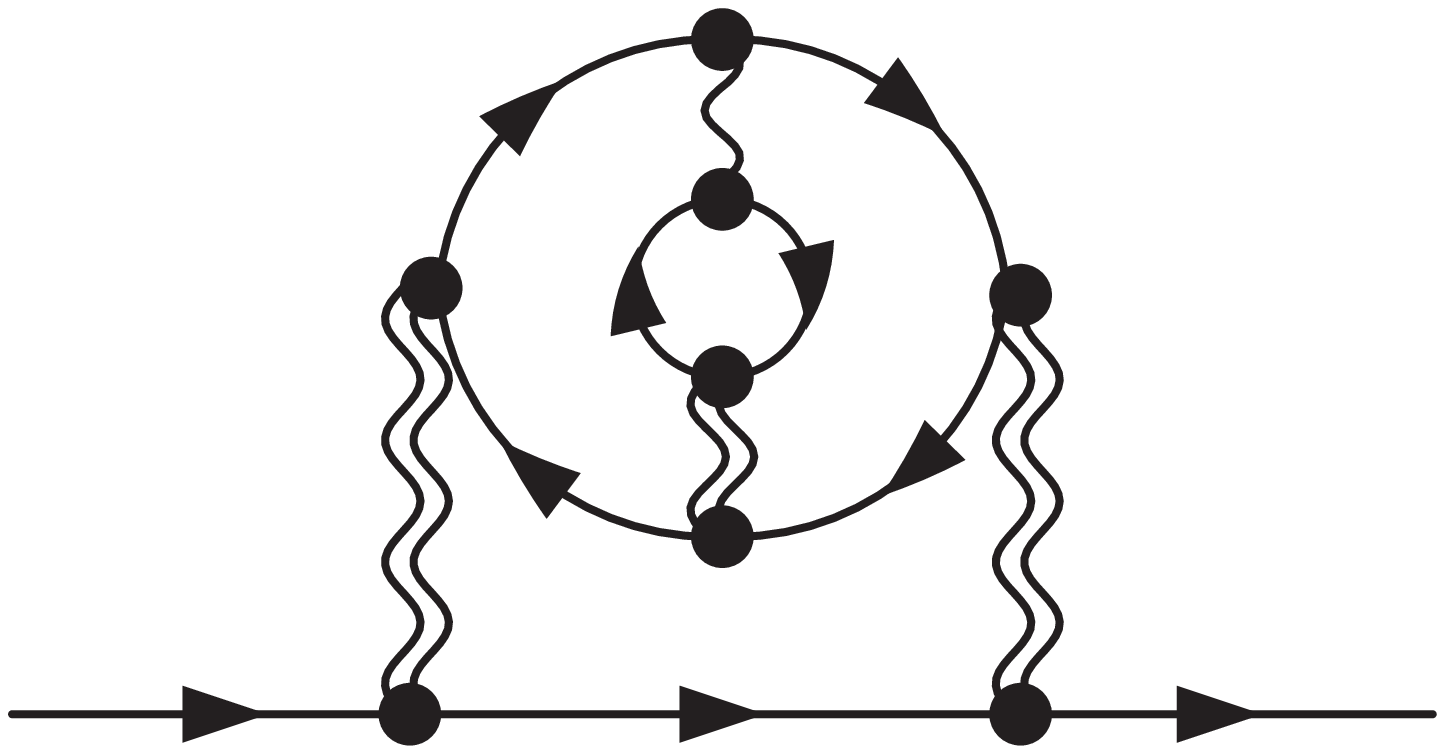}
 \end{minipage}
 \caption{
The order $1/(N-1)^2$ self-energy  at half-filling.
}
\label{fig:sg_rpa_n2}
\end{figure}

Higher-order fluctuations beyond the RPA 
appear first through the next-leading-order contributions. 
Figures \ref{fig:vertex_rpa} and \ref{fig:sg_rpa_n2}
show the order $1/(N-1)^2$ diagrams 
for the vertex function and the self-energy  
in the particle-hole symmetric case, respectively. 
These contributions and the higher-order components 
from Eq.\ \eqref{eq:U_bub_def} 
determine the renormalization factor
 $z$  and the Wilson ratio $R$ to order $1/(N-1)^2$. 
In the present work, we have calculated all these 
next-leading-order contributions. 
An outline of the calculations 
is given in a previous paper \cite{Ao2012}. 
We have also carried out the NRG calculations for $N=4$ 
to demonstrate the reliability of the $1/(N-1)$ expansion.

%

\begin{figure}[t]
\leavevmode
\begin{minipage}[t]{0.7\linewidth}
\includegraphics[width=\linewidth]{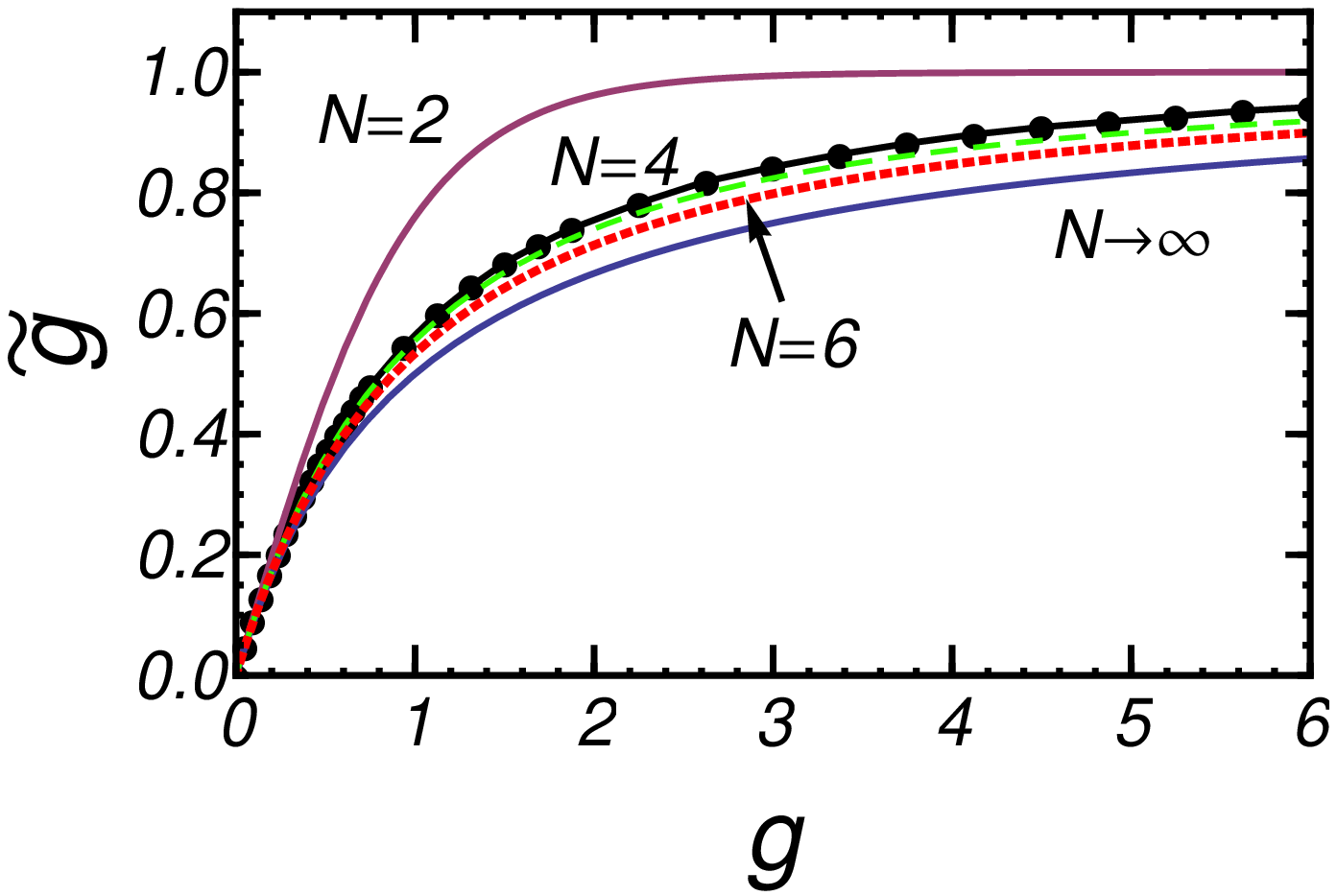}
\includegraphics[width=\linewidth]{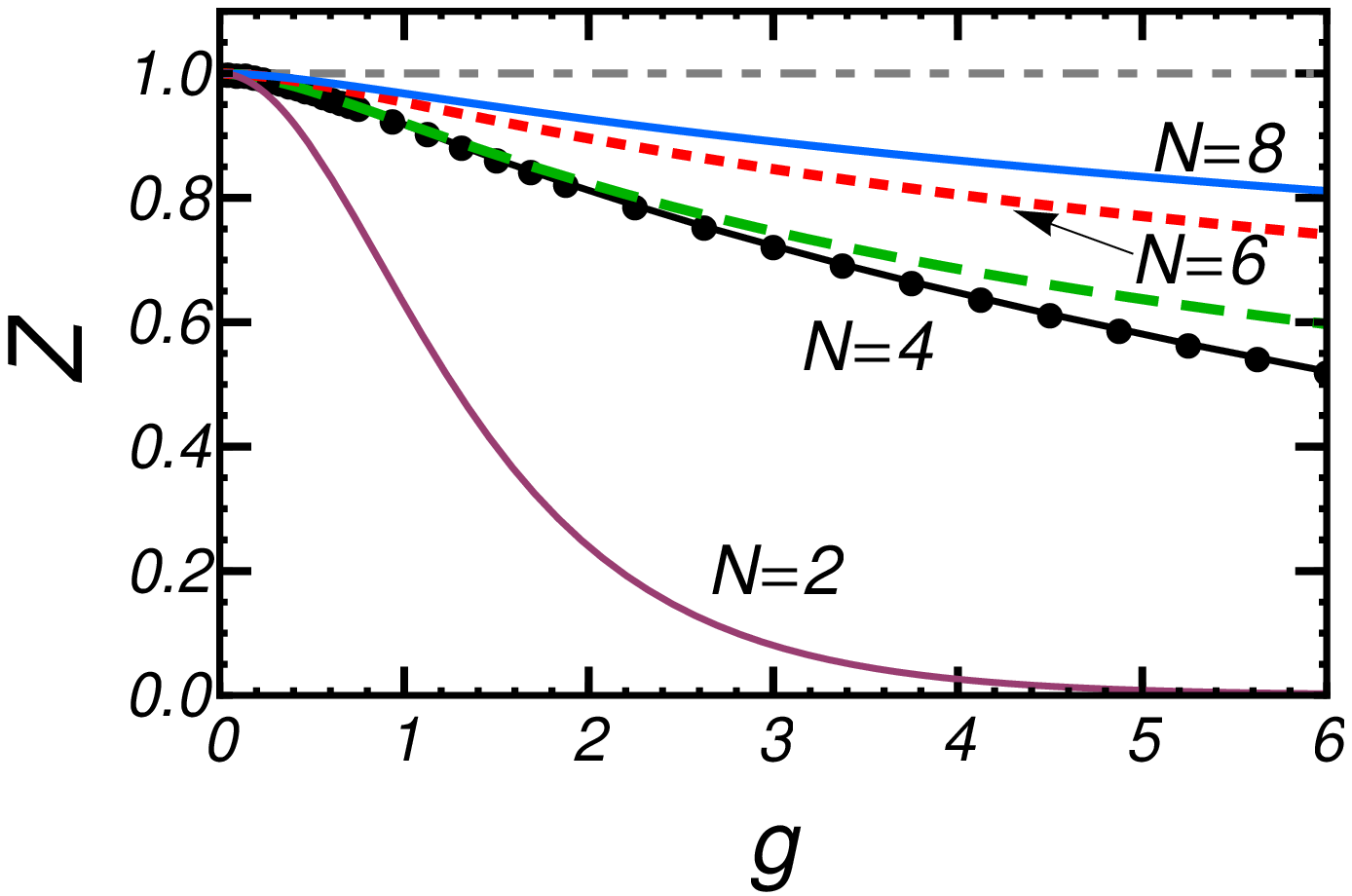}
\end{minipage}
\caption{
(Color online) 
Renormalized parameters  $\widetilde{g}$  and $z$
in the particle-hole symmetric case 
 $\epsilon_d=-(N-1)U/2$ 
are plotted 
as  functions of the scaled Coulomb interaction $g$ for $N=2$ 
(Bethe ansatz \cite{ZlaticHorvatic}), 
and for $N=4,6,8$  
(next-leading order in the $1/(N-1)$ expansion). 
 For $N=4$, the NRG results \cite{Nishikawa1,Sakano} 
are also shown with the solid circles ($\bullet$).
In the $N \to \infty$ limit,  $\widetilde{g}$  and $z$ approach 
 $\widetilde{g} \to g/(1+g)$ and $z \to 1.0$. 
}
\label{fig:g*_g}
\end{figure}

Figure \ref{fig:g*_g} shows  the next-leading-order 
results for 
$\widetilde{g}$ and $z$, plotted as functions of $g$.
We see the very close agreement
between the NRG and the next-leading-order results in the $1/(N-1)$ expansion
for $N=4$, especially for the renormalized coupling $\widetilde{g}$.
The two curves for $\widetilde{g}$, for $N=4$,   
almost overlap each other over the whole range of $g$ 
although the next-leading-order results (green dashed line) are 
 slightly smaller than those of the NRG (solid circles). 
As $N$ increases,  $\widetilde{g}$  converges 
rapidly to the RPA value, $\widetilde{g} \to g/(1+g)$, 
which is asymptotically exact in the limit of $N\to\infty$.
We also see that the value that $\widetilde{g}$ can take  
is bounded in a very narrow region  
between the curve for $N=4$ and that for the $N\to \infty$ limit.

The order $1/(N-1)^2$ results for the renormalization factor $z$,
shown in Fig.\ \ref{fig:g*_g}, 
also agree well with the NRG results for $N=4$ 
at $g \lesssim 3.0$, or equivalently 
in the region where $\widetilde{g} \lesssim 0.8$. 
This indicates that 
the next-leading-order results also reproduce  
the Kondo energy scale, $\widetilde{\Delta} = z \Delta$,  
properly from the weak-coupling to intermediate-coupling regions   
where $\widetilde{g}$ is still not very close to $1.0$, 
the value in the strong coupling limit.

\section{Full-counting statistics}

The $1/(N-1)$ expansion can be applied 
to nonequilibrium transport at finite bias voltages\ $V$.
To be specific, 
we choose the lead-dot couplings and chemical potentials 
in the leads to be symmetric, $\Gamma_L = \Gamma_R$  and  
$\mu_{L}= -\mu_{R}$ ($= eV/2$), and consider 
a steady state in the particle-hole symmetric case. 
In this case, the retarded Green's function, 
which is asymptotically exact 
at low energies  up to terms of order $\omega^2$, $T^2$, and $(eV)^2$, 
is given by \cite{ao2001,Sakano} 
\begin{align}
& 
\!\!\!\!\!\!\!\!\!\!\!\!\!\ 
G^r(\omega) 
 \simeq  
\frac{z}{ 
\omega + i \widetilde{\Delta} 
+i
\frac{{\widetilde{g}}^2}{2(N-1)\widetilde{\Delta}} 
\left[ \omega^2 + \frac{3}{4}(eV)^2 + (\pi T)^2 \right] 
}.
\end{align}
The average and the fluctuations of the  steady 
current  at low energies can be  described 
by the local Fermi-liquid theory, which follows from this form 
of the Green's function. 
 The Fermi-liquid theory, 
or the related renormalized perturbation theory (RPT) \cite{Hewson}, 
can also be applied to the full-counting statistics 
for nonequilibrium transport. 
Specifically, we consider the probability distribution $P({\bm q})$ 
of the transferred charge ${\bm q} = (q_1^{}, q_2^{}, \cdots, q_N^{})$
from the left lead to  the orbital $m$ of the Anderson impurity 
during a time interval ${\cal T}$.
The generating function for this probability distribution 
is defined by 
$\chi \left( {\bm \lambda} \right) 
= \sum_{\bm q} e^{i {\bm \lambda} \cdot {\bm q}}  P({\bm q})$
with the counting fields 
${\bm \lambda}=(\lambda_1^{},\lambda_2^{}, \cdots, \lambda_N^{})$,
and can be expressed in the form \cite{LevitovReznikov}
\begin{eqnarray}
\!\!\!\!\!\!\!\!\!\!\!
\chi \left( {\bm \lambda} \right)
\,=\, 
\left\langle 
 T_{C} \exp \left\{ -i \! \int_{C} \!dt \,
\bigl[ {\cal H}_T^{\lambda}(t) + {\cal H}_U^{}(t) \bigr] \right\}
\right \rangle .
\label{eq:CGF}
\end{eqnarray}
Here, $T_C$ is 
the time-ordering operator along the Keldysh contour $C$, 
and the time evolution 
is defined with respect to an extended Hamiltonian 
${\cal H}^{\lambda}(t) 
= {\cal H}_0^{} + {\cal H}_T^{\lambda}(t)+{\cal H}_U^{}$: 
\begin{align}
&
\!\!\!\!\!\!\!\!\!\!\!\!\!
{\cal H}_T^{\lambda}(t) 
= \sum_{k m} \left[ v_L^{} e^{i\lambda_m^{}(t)/2} 
d_m^{\dagger} c_{k L m}^{} 
+
v_R^{}  d_m^{\dagger}  c_{k R m}^{}
\right]
+ \mbox{H.c.} 
\end{align}
Here, the counting field depends on the path 
such that $\lambda_m^{} (t_{\mp}) = \pm \lambda_m^{}$ 
for the forward ($t_{-}$) and backward ($t_{+}$) paths,  respectively, 
and is switched on during a time interval $[t: 0 \to {\cal T} \to 0]$.

We obtained an explicit expression of $\chi(\bm{\lambda})$, 
which is asymptotically exact at low energies up to order $(eV)^3$ 
for general $N$ and $U$, 
using the RPT  
\cite{SakanoFCS,SakanoFCS2}.
The cumulant for the full current can be derived from $\chi(\bm{\lambda})$,
choosing the counting fields to be $\lambda_m^{}=\lambda$ for all $m$  
and then taking a derivative  
${\cal C}_n^{} = (-i)^n\frac{d_{}^n}{d\lambda_{}^n}\ln \chi(\lambda)$: 
\begin{align}
& 
\!\!\!\!\!\!\!\!\!\!\!\! 
\frac{\mathcal{C}_n^{}}{\mathcal{T}}
= \frac{J_u}{e} \!
\left[  \delta_{1n} 
+ \frac{(-1)^n}{12}  
\left\{1 + \frac{\left(1+2^{n+1}\right) \widetilde{g}^2}{(N-1)}\right\}
\!\left(\frac{eV}{\widetilde{\Delta}}\right)^2 
\right]. 
\label{eq:FullCCumulant}
\end{align}
Here, $J_u = Ne^2V/(2 \pi \hbar)$ is the linear-response current 
of the unitary limit, 
and $\delta_{nn'}$ is the Kronecker's delta. 
 Equation\ \eqref{eq:FullCCumulant} shows that the nonequilibrium properties 
can also be characterized by the two renormalized parameters, 
$\widetilde{g}$  and 
the Kondo energy scale $\widetilde{\Delta}$, 
 at low energies in the particle-hole symmetric case. 
For $n=1$, the cumulant $e\, {\cal C}_1^{}/{\cal T}$ expresses 
the steady current.
%
The universal properties of the cumulants 
for $n \geq 2$ can be extracted 
from the Fano-factor-{\it inspired\/} ratio (FFIR), which is 
normalized with respect to the backscattering current 
$J_b \equiv J_u - e \,{\cal C}_1^{}/{\cal T}$ \cite{GogolinKomnikPRL}:  
\begin{align}
& 
\!\!\!\!\!\!\!\!\!\!\!\! 
\frac{{\cal C}_n^{}}{{\cal C}_n^P}\,
=\, 
\frac{\displaystyle 1+\frac{\left(1+2^{n+1}\right)\,
\widetilde{g}^2}{N-1}}{\displaystyle 1 + \frac{5\widetilde{g}^2}{N-1}},
\qquad \  \frac{{\cal C}_{n}^P}{\mathcal{T}} \,\equiv\,
 (-1)^n \frac{J_b^{}}{e} .  
\label{eq:FFIR}
\end{align}
Here, ${\cal C}_{n}^P$ can also be regarded as 
the Poisson value of the cumulant, 
and the FFIR for $n=2$  corresponds to the Fano factor \cite{Sakano}.

\begin{figure}[tb]
\begin{minipage}[t]{0.60\linewidth}
\includegraphics[width=\linewidth]{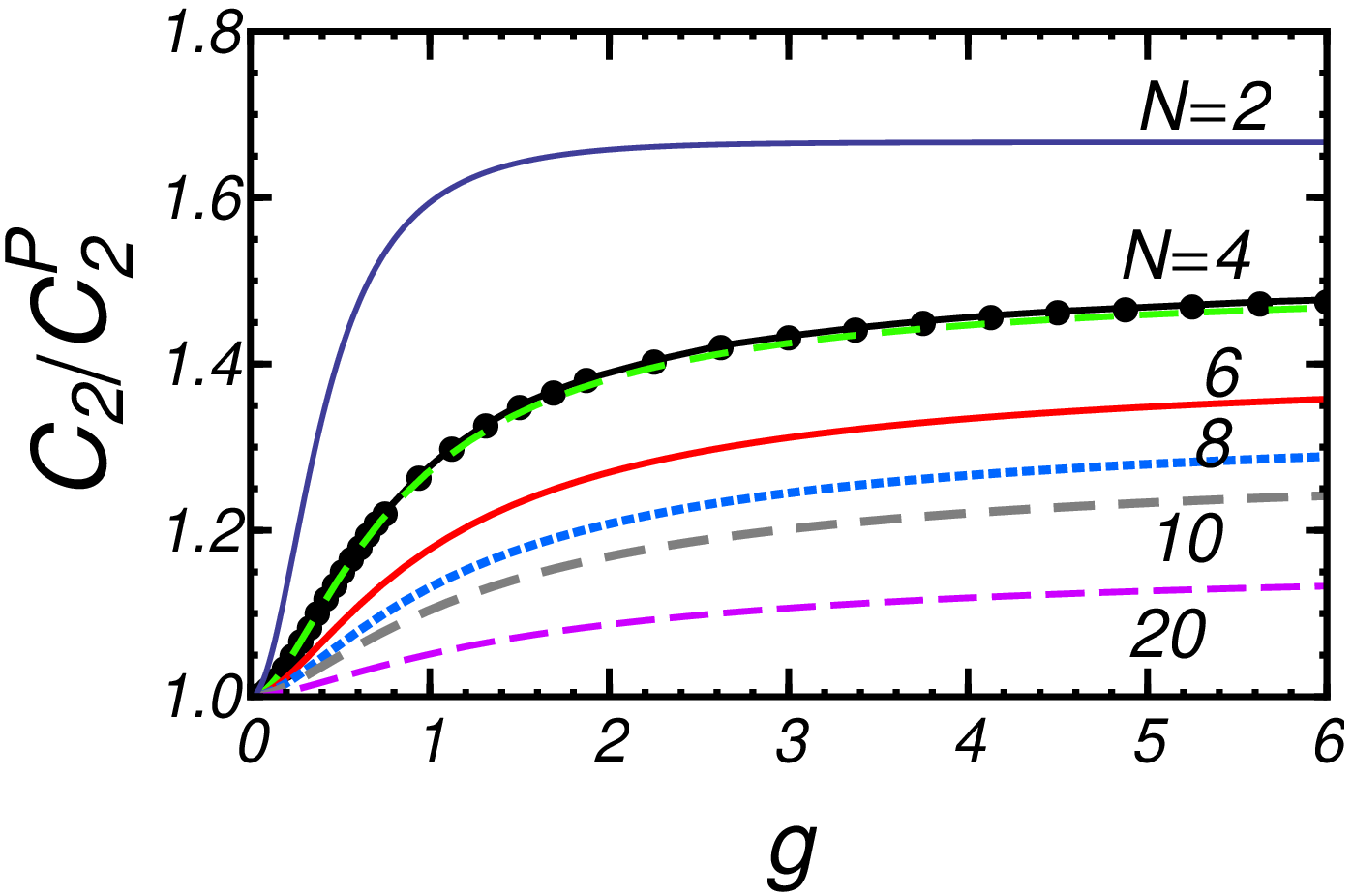}
\end{minipage}\\
\begin{minipage}[t]{0.60\linewidth}
\includegraphics[width=\linewidth]{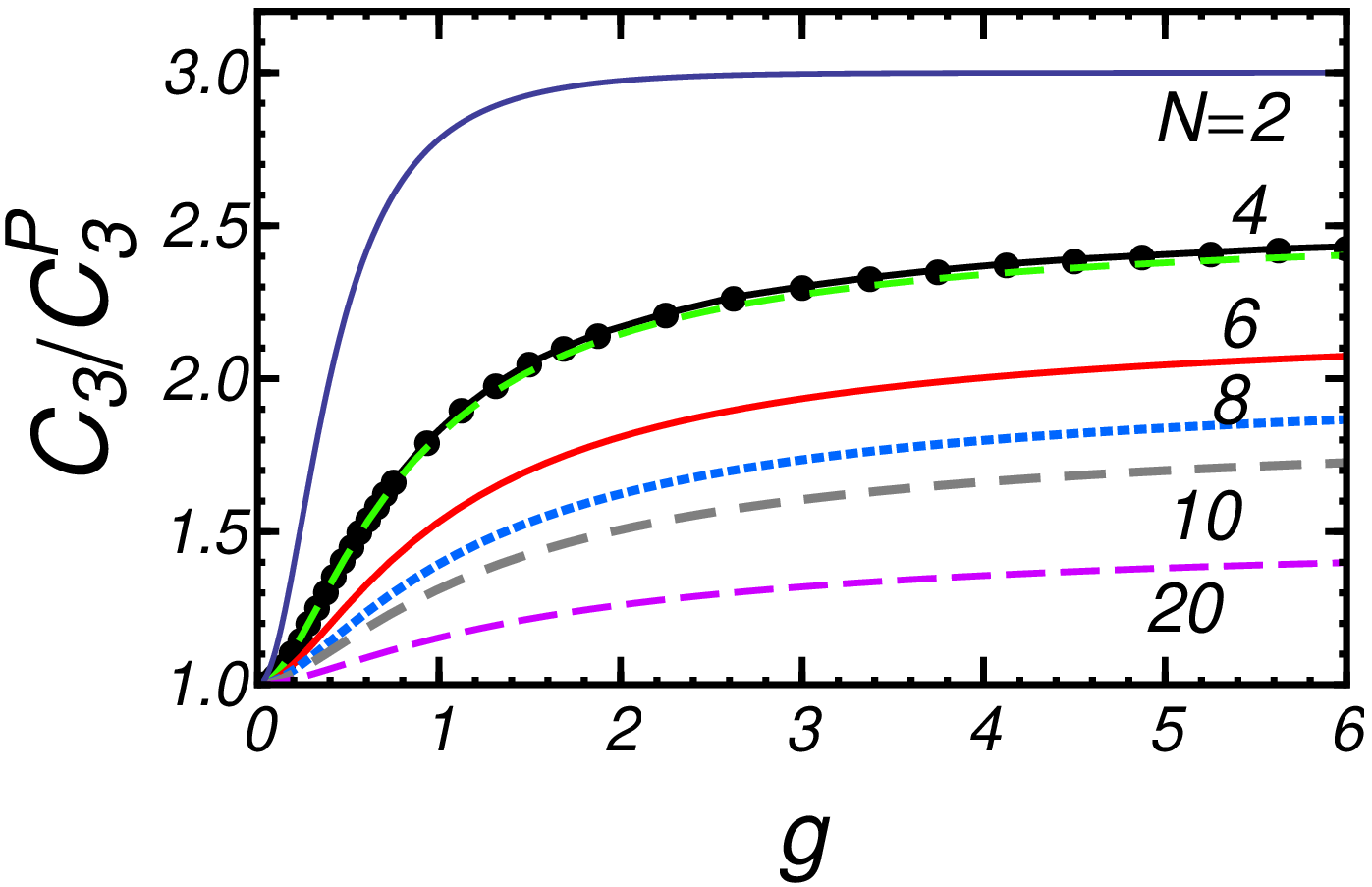}
\end{minipage}\\
\rule{0.16cm}{0cm}
\begin{minipage}[t]{0.58\linewidth}
\includegraphics[width=\linewidth]{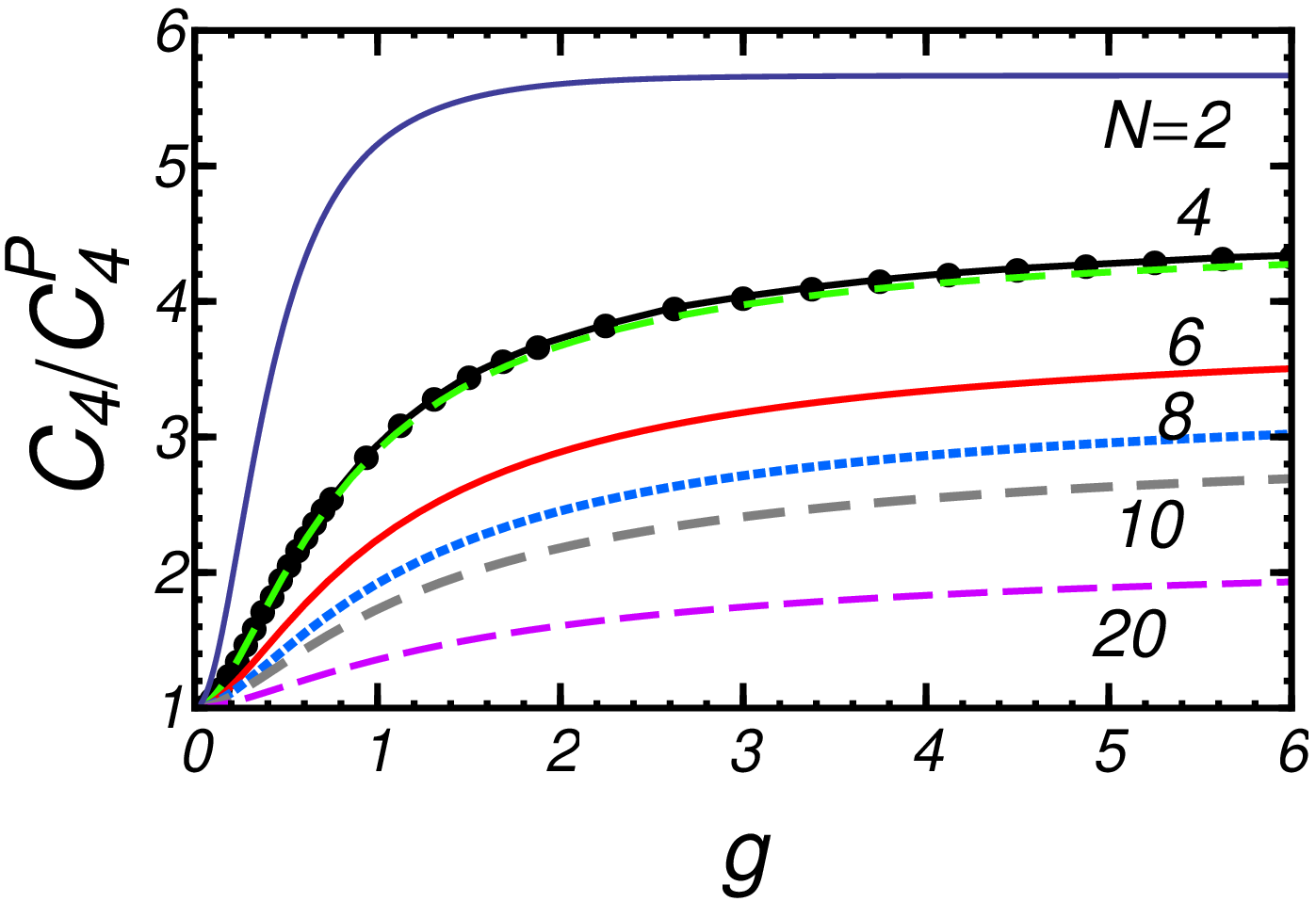}
\end{minipage}
\caption{
(Color online)
The FFIR ${\cal C}_n^{}/{\cal C}_n^P$ for 
$n=2$ (top), $n=3$ (middle), and n=$4$ (bottom) 
are plotted as  functions of $g$ for $N=2$ (Bethe ansatz), and 
$N=4,6,8,10, 20$ (next-leading-order results in the $1/(N-1)$ expansion).
The NRG results (solid circles) are also shown for $N=4$.}
\label{fig:FFIR}
\end{figure}

Figure \ref{fig:FFIR} shows 
the ratios ${{\cal C}_n^{}}/{{\cal C}_n^P}$ 
for the noise $n=2$ (top),  the skewness $n=3$ (middle), and 
sharpness $n=4$ (bottom) as functions of the scaled Coulomb
interaction $g$ for several value of the orbital degeneracy $N$.
We see close agreement between the NRG and 
the next-leading-order results for $N=4$ again. 
As the next-leading-order $1/(N-1)$ results 
for $\widetilde{g}$ are numerically almost 
exact for $N\geq 4$, the results shown in Fig.\ \ref{fig:FFIR}
capture orbital effects correctly.
In the strong coupling limit $g \to \infty$,
the FFIR approaches  ${{\cal C}_n^{}}/{{\cal C}_n^P} \to (N+2^{n+1})/(N+4)$ 
as the renormalized coupling converges to $\widetilde{g} \to 1$. 
As $N$ increases, $\widetilde{g}$ converges rapidly to 
the value of $\widetilde{g} \simeq g/(1+g)$ as mentioned above.
Thus, for $N \gtrsim 8$, the $N$ dependence is determined essentially 
by the factor $1/(N-1)$  appearing explicitly 
in Eq.\ \eqref{eq:FFIR}.
In the limit of $N \to \infty$, the cumulant approaches the Poisson value 
 ${\cal C}_n^{}/{\cal C}_n^P \to 1$  
as the fluctuations due to  eletctron correlations are suppressed and 
the mean-field theory becomes asymptotically exact.

In summary, carrying out the $1/(N-1)$ expansion  
 up to order $1/(N-1)^2$, 
we have obtained the cumulants for the probability 
distribution of the current through 
a quantum dot with orbital degeneracy $N>4$ at low energies. 
The results of the renormalized coupling $\,\widetilde{g}\,$ 
show excellent agreement  at $N = 4$ with the exact NRG results  
in the particle-hole symmetric case. 
This enable us to obtain 
{\em almost\/} exact numerical results for   
the Fano-factor-inspired ratio 
${\cal C}_n^{}/{\cal C}_n^P$  for $N>4$.

\begin{acknowledgments}
The authors thank T.\ Fujii, T.\ Kato, and A.\ C.\ Hewson for discussions. 
This work is supported by JSPS Grants-in-Aid for 
Scientific Research C (No.\ 23540375, and No.\ 24540316).
Numerical computation was partly carried out 
at the Yukawa Institute Computer Facility.
\end{acknowledgments}


\begin{references}


\bibitem{Grobis}
M.\ Grobis, 
I.\ G.\ Rau, R.\ M.\ Potok, H.\ Shtrikman, 
and D.\ Goldhaber-Gordon,
Phys.\ Rev.\ Lett.\ {\bf 100}, 246601 (2008).


\bibitem{ScottNatelson}
G.\ D.\ Scott, 
 Z.\ K.\ Keane, J.\ W.\ Ciszek, J.\ M.\ Tour, and D.\ Natelson, 
Phys.\ Rev.\ B {\bf 79}, 165413 (2009).


\bibitem{KNG}
A.\ Kaminski, Yu.\ V.\ Nazarov, and L.\ I.\ Glazman,
  Phys.\ Rev.\ B {\bf 62}, 8154 (2000).


\bibitem{ao2001}
A.\ Oguri, Phys.\ Rev.\ B {\bf 64}, 153305 (2001).


\bibitem{FujiiUeda} 
T.\ Fujii and K.\ Ueda, 
Phys.\ Rev.\ B {\bf 68}, 155310  (2003).


\bibitem{HBA} 
A.\ C.\ Hewson, J.\ Bauer, and A.\ Oguri,
J.\ Phys.: Condens.\ Matter {\bf 17}, 5413 (2005).




\bibitem{Heiblum} 
O.\ Zarchin, M.\  Zaffalon,  M.\ Heiblum, D.\ Mahalu, and V.\ Umansky, 
Phys.\ Rev.\ B {\bf 77}, 241303  (2008).


\bibitem{Kobayashi}
Y.\ Yamauchi,  
 K.\ Sekiguchi, K.\ Chida, T.\ Arakawa, S.\ Nakamura, 
K.\ Kobayashi, T.\ Ono, Teruo, T.\ Fujii, and R.\ Sakano, 
Phys.\ Rev.\ Lett.\ {\bf 106}, 17660 (2011).


\bibitem{GogolinKomnik} 
A.\ O.\ Gogolin and A.\ Komnik, 
Phys.\ Rev.\ B {\bf 73}, 195301 (2006).


\bibitem{GogolinKomnikPRL}
A.\ O.\ Gogolin, and  A.\ Komnik,
Phys.\ Rev.\ Lett.\ {\bf 97}, 016602 (2006).

\bibitem{Golub} 
A.\ Golub, Phys.\ Rev.\ B {\bf 73}, 233310 (2006).


\bibitem{Sela2009} 
E.\ Sela and J.\ Malecki,
Phys.\ Rev.\ B {\bf 80}, 233103 (2009). 


\bibitem{Fujii2010}
T.\ Fujii, J.\ Phys.\ Soc.\ Jpn.\ {\bf 79}, 044714 (2010). 



\bibitem{Delattret} 
T.\ Delattre,  
 C.\ Feuillet-Palma, L.\ G.\ Herrmann, P.\ Morfin, 
 J.-M.\ Berroir, G.\ F\`{e}ve, 
 B.\ Pla\c{c}ais, D.\ C.\ Glattli, M.-S. Choi, C. Mora,
 and T. Kontos,
 Nature Phys.\ {\bf 5}, 208 (2009).


\bibitem{Mora2009}
C.\ Mora, P.\ Vitushinsky, X.\ Leyronas, A.\ A.\ Clerk, and
K.\ Le Hur, Phys. Rev. B {\bf 80}, 155322 (2009).


\bibitem{Sakano}
R.\ Sakano, T.\ Fujii, and A.\ Oguri, 
Phys.\ Rev.\ B {\bf 83}, 075440  (2011).


\bibitem{SakanoFCS}
 R.\ Sakano, A.\ Oguri, T.\ Kato and S.\ Tarucha,
 Phys.\ Rev.\ B {\bf 83}, 241301 (2011).


\bibitem{SakanoFCS2}
 R.\ Sakano,  Y.\ Nishikawa, A.\ Oguri, A.\ C.\ Hewson, and S.\ Tarucha,
 Phys.\ Rev.\ Lett.\ 
{\bf 108}, 266401 (2012). 


 \bibitem{KWW}
  H.\ R.\ Krishna-murthy, J.\ W.\ Wilkins, and K.\ G.\ Wilson, 
  Phys.\ Rev.\ B {\bf 21}, 1003 (1980). 


\bibitem{Izumida}
W.\ Izumida, O.\ Sakai and Y.\ Shimizu,
J.\ Phys.\ Soc.\ Jpn.\ {\bf 67}, 2444  (1998).




\bibitem{Coleman} 
P.\ Coleman, 
Phys.\ Rev.\ B {\bf 28}, 5255 (1983).


\bibitem{Bickers}
N.\ Bickers, Rev.\ Mod.\ Phys.\ {\bf 59}, 845 (1987).



\bibitem{AoSakanoFujii}
 A.\ Oguri,  R.\ Sakano, and T.\ Fujii,  
Phys.\ Rev.\ B {\bf 84}, 113301 (2011).


\bibitem{Ao2012}
 A.\ Oguri,    
Phys.\ Rev.\ B {\bf 85}, 155404 (2012).


\bibitem{LevitovReznikov}
L.\ S.\ Levitov  and M.\ Reznikov,
Phys.\ Rev.\ B {\bf 70}, 115305 (2004).


\bibitem{Bagrets}
D.\ Bagrets, Y.\ Utsumi, D.\ Golubev, and G.\ Sch\"{o}n,
Fortschr.\ Phys.\ {\bf 54}, 917 (2006).

\bibitem{Esposito}
M.\ Esposito, U.\ Harbola, and S.\ Mukamel, 
Rev.\ Mod.\ Phys.\ {\bf 81}, 1665 (2009).


\bibitem{YY2}
  K. Yamada, 
  Prog.\ Theor.\ Phys. {\bf 53}, 970 (1975).


\bibitem{Yoshimori}
 A.\ Yoshimori, Prog.\ Theor.\ Phys.\ {\bf 55}, 67 (1976).


\bibitem{ZlaticHorvatic}
V.\ Zlati\'c and B.\ Horvati\'c,
Phys.\ Rev.\ B {\bf 28}, 6904 (1983).

 \bibitem{Nishikawa1}
 Y.\ Nishikawa, D.\ J.\ G.\ Crow, and A.\ C.\ Hewson, 
 Phys.\ Rev.\ B {\bf 82}, 115123 (2010).

\bibitem{Hewson}
 A.\ C.\ Hewson, 
 J.\ Phys.:\ Condens.\ Matter {\bf 13}, 10011 (2001).





\end{references}
\end{document}